\documentclass[aps,prx,10pt,twocolumn,showpacs,superscriptaddress,floatfix,longbibliography]{revtex4-2}
\usepackage[english]{babel}
\usepackage[utf8]{inputenc}
\usepackage{graphicx}
\usepackage[unicode=true,pdfusetitle,
 bookmarks=true,bookmarksnumbered=false,bookmarksopen=false,
 breaklinks=false,pdfborder={0 0 0},pdfborderstyle={},backref=false,colorlinks=true]
 {hyperref}
\hypersetup{
 citecolor=blue,
 urlcolor=blue,
 linkcolor=blue
}
\usepackage{latexsym,amsmath,verbatim}
\usepackage{amssymb} 
\usepackage{amsfonts}
\usepackage{colortbl}
\usepackage{array}
\usepackage{stackrel}
\usepackage{bm}
\usepackage{nicefrac}
\usepackage{rotating}
\usepackage{float}
\usepackage[final]{pdfpages}
\usepackage{lipsum}  
\usepackage{braket}
\usepackage{dsfont}
\usepackage[normalem]{ulem}
\usepackage{xcolor}
\usepackage{nameref}
\usepackage{cleveref}
\usepackage{csquotes}

\DeclareMathOperator{\Tr}{Tr}

\makeatletter
\AtBeginDocument{\let\LS@rot\@undefined}
\makeatother

\crefname{figure}{Fig.}{Figs.}
\crefname{section}{Sec.}{Secs.}

\begin{document}

\title{Tree networks of real-world data: analysis of efficiency and spatiotemporal scales}

\author{Davide Cipollini}
    \email{d.cipollini@rug.nl}
    \affiliation{Bernoulli Institute for Mathematics, Computer Science and Artificial Intelligence, University of Groningen, Netherlands}
    \affiliation{Cognigron - Groningen Cognitive Systems and Materials Center, University of Groningen, Netherlands}
\author{Lambert Schomaker}
    \affiliation{Bernoulli Institute for Mathematics, Computer Science and Artificial Intelligence, University of Groningen, Netherlands}
    \affiliation{Cognigron - Groningen Cognitive Systems and Materials Center, University of Groningen, Netherlands}

\begin{abstract}
Hierarchical tree structures are common in many real-world systems, from tree roots and branches to neuronal dendrites and biologically inspired artificial neural networks, as well as in technological networks for organizing and searching complex datasets of high-dimensional patterns.
Within the class of hierarchical self-organized systems, we investigate the interplay of structure and function, associated with the emergence of complex tree structures in disordered environments. 
Using an algorithm that creates and searches trees of real-world patterns, our work stands at the intersection of statistical physics, machine learning, and network theory. We resolve the network properties over multiple phase transitions and across a continuity of scales, using the von Neumann entropy, its generalized susceptibility, and the recent definition of thermodynamic-like quantities, such as work, heat, and efficiency.
We show that scale-invariance, i.e. power-law Laplacian spectral density, is a key feature to construct trees capable of combining fast information flow and sufficiently rich internal representation of information, enabling the system to achieve its functional task efficiently.
Moreover, the complexity of the environmental conditions the system has adapted to is encoded in the value of the exponent of the power-law spectral density, inherently related to the network spectral dimension, and directly influencing the traits of those functionally efficient networks.
Thereby, we provide a novel metric to estimate the complexity of high-dimensional datasets.


\end{abstract}


\maketitle

\section{Introduction}
\label{sec:intro}
Why complex structures emerge in a real-world environment is a fascinating question that has not yet found a fully definite answer. 
In a recent article~\cite{Ghavasieh2024}, the emergence of complex structures is devised as the macroscopic thermodynamic result of, on the one side, the optimal ability to store information about the environment in an accessible manner, on the other side, the ability to transmit information across the structure~\citep{Szathmry1995} (see Methods \Cref{sec:thermodynamics_of_netw}).
For instance, in living systems, information about the environment flows into the system's genome through natural selection. The genome encodes such information into the organism's structural traits, which in turn reflect the functional capability of the system. Thus, the system's very own structure may be regarded as the internal representation of the environment~\citep{Hopfield1982,Ackley1985,Kachman2017,perunov2016statistical,Domenico_2024} and its
the structural traits could themselves enclose the complexity of the environmental conditions the system needs to adapt to.
How these traits are optimized to maximize the fitness to the environment is an open question. The hypothesis of optimal functioning at the border of order and disorder suggests critical-like features as the potential candidates to obtain functional advantages~\citep{Langton1990,Bertschinger2004,Mastromatteo2011,Hidalgo2014,Daniels_2018,Krotov2014,Beggs2004} and maximize the fitness to the outer world, or, when in the context of learning machines, to obtain a better performance in a statistical generalization problem~\cite{Mastromatteo2011}. Among the various attributes exhibited by truly critical systems and systems with critical-like properties, especially relevant to our work is the property of scale-invariance, exhibited through the power-law distribution of some meaningful quantity of the system under consideration.

\begin{figure}[ht!]
        \centering
		\includegraphics[width=1\columnwidth]{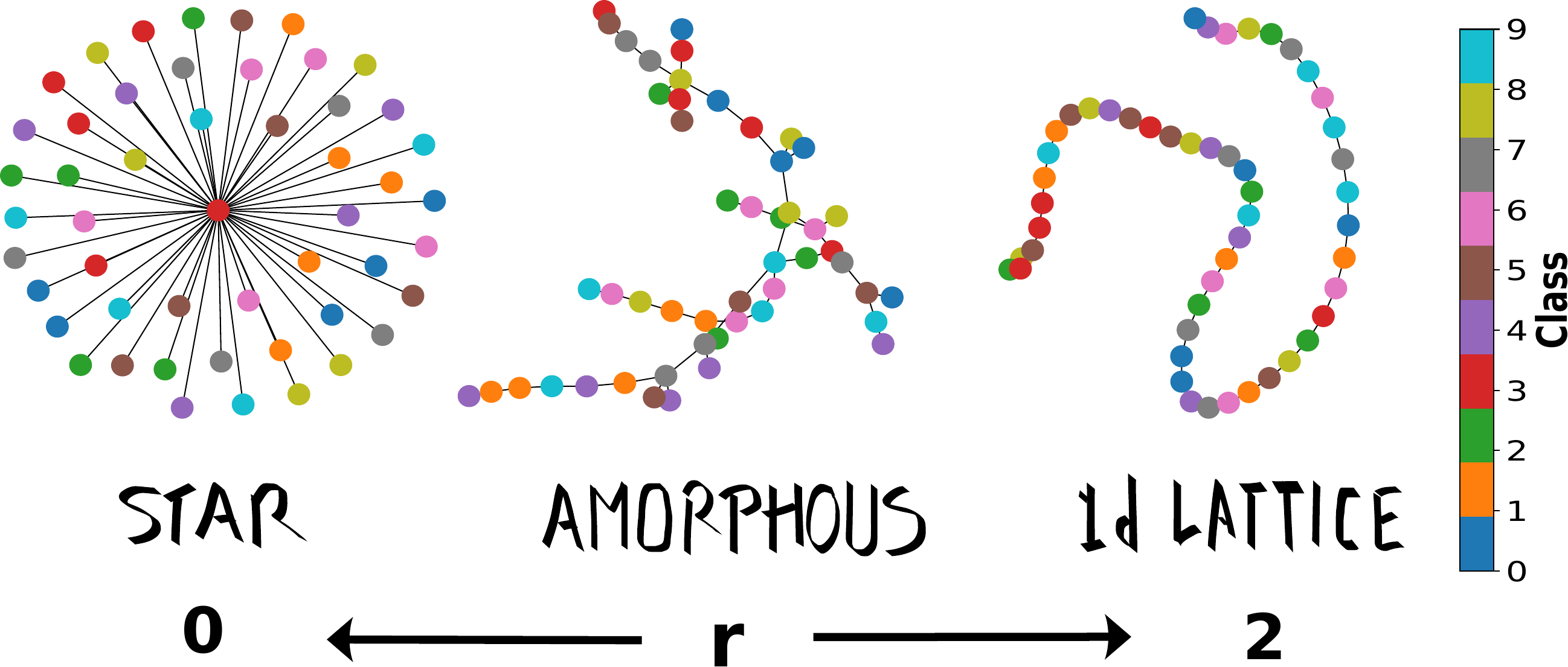}
	\caption[]{\textbf{Tree structures generated by the Portegys algorithm.} The algorithm creates star-like networks up to 1d lattice when $r\rightarrow2$. For intermediate values of $r$, networks resembling an arborescent tree are produced. The process is reminiscent of the transition \textit{vapor/liquid} $\rightarrow$ \textit{amorphous solid} $\rightarrow$ \textit{1d lattice}. Depicted networks are generated from a subset of 50 samples randomly extracted from the MNIST dataset with $r=0.05, 0.95, 1.50$ respectively. Classes are indicated by the 10 different colors in the color bar.}
	\label{fig:structures}
\end{figure}

Hierarchical trees enable us to investigate the interplay between structure and function, associated with the emergence of complex structures in disordered environments, within a relevant class of systems.
Trees are common to many real-world systems, from tree roots and branches, to neuronal dendrites~\cite{Shepherd1987,Poirazi2020,Hodassman2022,Kastellakis2016,Makarov2023} and biologically inspired artificial neural networks~\cite{Chavlis2025,Guerguiev2017,Payeur2021,Sacramento2018,Li2020,Lee2015}, as well as technological networks for organizing and searching collections of high-dimensional patterns~\cite{Nielsen2016,Portegys,Mangalagiu1999}.

We use the Portegys algorithm~\cite{Portegys,Mangalagiu1999} to produce hierarchical trees that self-organize given a collection of high-dimensional vectors.
The algorithm recursively organizes (structural aspect) high-dimensional vectors randomly drawn from a dataset into trees (see in Methods \Cref{sec:portegys_alg}). 
Using the metaphor introduced by Jacob for natural selection~\citep{Jacob1977}, the Portegys tree-generation process acts as a \textit{tinkerer} that creates a structure using what is available at the time, collocating one node after the other across the tree formation process according to a parameterized distance-based rule. 
The algorithm integrates a search phase (functional aspect) where it navigates the tree structure to achieve efficient pattern-matching.

The algorithm is simple, explainable, and subject to a single control parameter, $r$. Still, it is capable of generating a large variety of branching structures (\Cref{fig:structures}) and multiple phase transitions (\Cref{sec:phasediagram}).
Other models can generate hierarchies. Such as the model in Ref.~\cite{Benatti2022} where a single parameter controls the branching and produces structures in much analogy to the ones in \Cref{fig:structures}, or the Galton–Watson stochastic process originally proposed to model the extinction of family names ~\cite{Watson1875,Harris2012}. Nevertheless, the Portegys algorithm integrates the functional aspect given by the pattern-matching task and includes the disorder of the external dataset, two key attributes in our analyses.
Thanks to these features, we can work at the intersection of three disciplines: statistical physics, machine learning, and network science. We can show that information-theoretical tools such as the von Neumann entropy~\cite{De_Domenico_2016} and its generalized susceptibility~\citep{Villegas_2022,Villegas2023} (see Methods), while monitoring the structural properties of the system, reflect the functional capabilities of the network.

We show that emerging efficient structures exhibit a scale-invariant spectrum~\cite{poggialini2024networks}, i.e. power-law Laplacian spectral density. We reproduce this result over multiple datasets, ranked based on their complexity.
Our findings suggest that a large range of availability in the power law spectral density provides the advantage of a large repertoire of network ``frequencies''. These can be exploited to achieve the functional purpose of the system efficiently, akin to critical-like systems exhibiting scale-invariance in the distribution of a relevant quantity~\cite{Kardar2007,Hidalgo2014,diSanto2018,Krotov2014,Beggs2004}. 
Such behavior is observed consistently whether we take efficiency as gauged by the task-agnostic network thermodynamic efficiency~\cite{Ghavasieh2024}, estimating the trade-off between the diversity of information pathways and fast information propagation (see \Cref{sec:thermodynamics_of_netw} in Methods), or the pattern-matching efficiency (\Cref{sec:search_phase} in Methods), quantifying the trade-off between accuracy and number of visited nodes within the pattern-matching task.

Furthermore, the exponent of the power-law spectral density, related to the network generalized dimension~\cite{Alexander1982,hattori1987gaussian,Burioni1996,Burioni2004}, results to be driven by the complexity of the environment the system adapted to, directly influencing the traits of those functionally efficient networks.

Finally, from a different perspective, using a hierarchical topology to map the distribution of high-dimensional vectors of a dataset provides insights into the structured disorder of the data~\cite{mezard2023spin}.
In particular, we measure the dataset network's spectral dimension and show that it estimates the dataset's complexity. Thus, among the other results, our work provides an alternative way to gauge the dataset's complexity~\cite{mezard2023spin,Mead2020,Lempel1976OnTC,levina2004maximum, Mackay_Ghahrami,pope2021intrinsic}.

\section{Methods}
\label{sec:methods}

\subsection{Theoretical framework}
\subsubsection{Statistical physics of networks}
\label{sec:statistical_phys}

Information spreading over networked structures is governed by the Laplacian matrix~\citep{Masuda_2017} and its spectrum of eigenvalues encodes many relevant topological properties of the graph~\citep{Anderson_1985,Estrada2011}. Let us consider an undirected and unweighted network $G(V, E)$ of order $|V|=N$ with the number of links equal to $|E|=m$. The network Laplacian is defined as $\hat{L}=\hat{D}-\hat{A}$, where $\hat{D}$ and $\hat{A}$ are the diagonal matrix of the node degrees and the adjacency matrix respectively.
The Laplacian operator can be used to define a density operator~\citep{De_Domenico_2016}, where the Laplacian takes the role of the Hamiltonian of statistical physics~\citep{Kardar2007} as in~\Cref{eq:rho}.

Given an initial state of the network encoding the concentration of a quantity, e.g. information, in one or more nodes, $\underline{s}(0)$, the state at time $\tau$ is given by $\underline{s}(\tau)=e^{-\tau\hat{L}}\underline{s}(0)$. Where $\hat{K}=e^{-\tau\hat{L}}$ is the network propagator and is the discrete counterpart of the path integral formulation for general diffusion processes~\citep{Feynman_1998,Moretti2019}.

As $\hat{K}$ is a positive semidefinite Hermitian matrix admits the spectral decomposition in terms of the sum of projectors $\hat{K}=\sum_i^N e^{-\tau\lambda_i}|k_i><k_i|$. Thus, for any state at $\tau$ it holds $|\phi(\tau)> = \sum_i^N e^{-\tau\lambda_i}|k_i><k_i||\phi (0)>$. This relation remarks that the eigenvectors of the Laplacian are akin to the wave vectors in the Fourier space defined by the spectrum of the continuum Laplacian. At the same time, $\tau$ can be interpreted as the ``wavelength'', $\ell \sim \tau$, of the diffusion modes, thus it defines an operational resolution scale for observations~\cite{Villegas2023}. Moreover, the ensemble of accessible diffusion modes at time $\tau$ can be constructed by defining a probability measure over the accessible modes decomposed by the spectral theorem. 
Such canonical ensemble~\citep{De_Domenico_2016, Ghavasieh2020, Villegas_2022} is encoded into the density matrix:
 \begin{equation}
\label{eq:rho}
\hat{\rho}(\tau)=\frac{e^{-\tau\hat{L}}}{Z}
\end{equation}
The elements $\rho_{ij}$ represent the normalized amount of information transferred in a diffusion process between nodes $i$ and $j$ at time $\tau$. Remarkably $\rho_{ij}$ takes into account all possible topological pathways between the two nodes and assigns smaller weights to longer ones tantamount to path-integrals~\citep{Feynman_1998,Villegas_2022}. The partition function $Z=\Tr[\hat{K}]=\sum^N_{i=1} e^{-\tau\lambda_i(\hat{L})}$ is a function of the eigenvalues, $\lambda_i$, of the Laplacian and is proportional to the average return probability of a continuous time edge-centric random walker to be back in its initial location after time $\tau$~\citep{Ghavasieh2020}. Note that the object $\hat{\rho}$ is a positive semi-definite and Hermitian matrix whose trace sums to unity allowing its eigenvalues to be interpreted as probabilities. 

Finally, the von Neumann spectral entropy can be introduced~\citep{De_Domenico_2016} as a multi-scale descriptor of the network structure:
\begin{equation}
    \label{eq:vonNeumann_entropy}
    S(\hat{\rho}(\tau))=-\sum_{i=1}^N \mu_i(\tau) \log \mu_i (\tau)
\end{equation}
where $\{\mu_i(\tau)\}_{i=1}^N$ is the set of $\hat{\rho}(\tau)$ eigenvalues that are related to the set of Laplacian eigenvalues through $\mu_i(\tau)=e^{-\tau \lambda_i}/Z$. The entropy $S(\hat{\rho}(\tau))$ is a function of the normalized time, $\tau$, and is bounded between $[\log C_{conn},~\log N ]$ where $C_{conn}$ is the number of connected components and $N$ is the number of nodes.
Thus the von Neumann entropy, $S$, as a function of $\tau$ reflects the \textit{entropic phase transition} of information propagation (or loss) over the network~\citep{Villegas_2022}. Specifically for a connected network, at $\tau \rightarrow 0$, $S(\tau)=\log N$ reflects the segregated heterogeneous phase where information diffuses from the single nodes to the very local neighborhood; at $\tau \rightarrow \infty$, $S(\tau)=0$ and the diffusion is governed by the smallest non-zero eigenvalue of the Laplacian, associated with the so-called Fiedler eigenvector, reflecting the integrated homogeneous phase where the information has propagated all over the network. 

The network entropy thus counts the (logarithmic) number of informationally unconnected nodes in the heterogeneous network. By the inspection of the entropy, $S$, and its generalized susceptibility (or specific heat), $C = - dS / d\log \tau$, it is possible to identify the characteristic resolution scales of the network~\citep{Villegas_2022,Villegas2023}. For instance, a single large peak of the specific heat, $C$, identifies a single scale at which there is a fast and unique information diffusion (or information loss) across the network. The latter is the typical situation of a star graph, depicted in~\Cref{fig:structures}, or an Erd\"{o}s-R\'enyi network. Before the peak, the informationally connected components are the single nodes, while after the peak the informationally connected components are the whole network, thus revealing a single trivial scale. On the contrary, a constant specific heat reflects the scale-invariance of the network in a resolution region where information is lost at a constant rate. Such a situation is typical of homogeneous $d$-dimensional lattices and of heterogeneous networks whose spectral density is power-law distributed, $\mathcal{P}(\lambda)\sim\lambda^{\gamma}$, in a sufficiently extended scale-interval~\cite{Villegas2023}. The latter sentence makes for a definition for the scale-invariant networks~\cite{poggialini2024networks}, characterized by a scale-invariant information propagation across the network resolution scales. The the specific heat plateau height, $C_0$, is also related to the spectral dimension, $d_s=2(\gamma+1)$~\cite{Alexander1982,Burioni1996}, according to the relation $d_s=2C_0$~\cite{Villegas2023,poggialini2024networks}, which implies $C_0=\gamma+1$.
The spectral dimension generalizes the concept of the Euclidean dimension to disordered structures and coincides with the Euclidean dimension for lattices. For instance, we expect $C_0=1/2$ for 1d-lattice; $C_0=1$ for the regular grid with periodic boundary conditions; and $C_0=4/6$ for random trees with finite degree variance $<k^2>$ showcasing fractal dimension $d_s=4/3$~\cite{poggialini2024networks}.

From these considerations, a Laplacian-Renormalization group theory has been recently developed, allowing for a coarse-graining of the network structures into super-nodes and for the unraveling of the mesoscale structures~\citep{Villegas2023, Villegas_2022}. The theory relies on information diffusion to identify \emph{informationally equivalent ``blocks''} akin to the ones proposed by Kadanoff~\cite{Kadanoff1966}, and moves to the $k$-space to explain the procedure akin to Wilson's formulation~\cite{Wilson1974}. In this context, the first peak of the specific heat, $C$, or a marked change in its slope, identifies the most convenient resolution scales where to perform the coarse-graining over the informationally connected nodes, as it allows the iteration of the procedure without an excessive network reduction, and, when the network is scale-invariant, maintaining the dilation symmetry across the renormalization group procedure~\cite{Kardar2007}.

To conclude, we can intuitively use the diffusion parameter, $\tau$, to resolve the network topological structures at different resolution scales. Thus, the necessity of analyzing complex networked systems at all levels is solved by resorting to the conjugate Fourier space, the \textit{k}-space, defined by the Laplacian spectrum and the ``wavelengths'' of its modes.

\subsubsection{Thermodynamics of networks}
\label{sec:thermodynamics_of_netw}
The statistical physics theory for networks introduced above allows us to view network phenomena from a new perspective and possibly to bridge the gap to network macroscopic observables. In this regard, network thermodynamic quantities such as work, $W$, and heat, $Q$, have been defined in mathematical analogy to their thermodynamic counterparts~\cite{Ghavasieh2024}.
Since a network of isolated nodes is the only network with maximum entropy $S_{iso}=\log N$ for $\tau>0$, progressively adding edges diminishes the network's entropy. Lower entropy reflects the diminished ability of the network to express diversity to a stochastic perturbation, or in other terms, the loss in the internal representation capabilities encoded in the network topological traits. Thus, the entropic cost is given by:
\begin{equation}
    \label{eq:heat}
    Q = - \frac{\delta S}{\tau} = -\frac{\log N - S}{\tau} \leq 0
\end{equation}
which mathematically resembles the heat dissipation in a thermodynamic process.

On the other hand, adding edges increases the gain in information flow quantified by the (Helmholtz) free energy $F=- \log Z/\tau$~\citep{Ghavasieh2024,Ghavasieh2020}. Less obvious is how, for a fixed number of edges, the degree distribution influences the flow of information. As the Laplacian governs the continuous-time \textit{edge-centric} random walk, the presence of large hubs implies an enhancement of the process with faster propagation of the diffusing quantity~\citep{Masuda_2017,Hens2019}.
As the free energy is inversely related to the partition function, $Z=\sum_{i=1}^N e^{-\tau \lambda_i}$, the bounds of the former are given by the bounds of the latter. From $1\leq Z \leq N$, where the lower bound is given for the connected network at large $\tau \gg 1$, while the upper bound is given by a network of isolated nodes $\forall \tau$, or for a connected one at small scales $\tau=0$.  
Thus, the $-\log N /\tau \leq F \leq 0$ and the gain in information flow is reflected by the variation of the free energy from the case of isolated nodes: 

\begin{equation}
    \label{eq:work}
    W = \delta F = F - F_{iso} = F + \frac{\log N}{\tau} \geq 0
\end{equation}
which is tantamount to the work in the canonical ensemble.
Moreover, $U= W + Q \geq 0$ which implies $W \geq |Q|$.
A formation process can be depicted as the trade-off between these two necessities of the networked complex system: the need to maximize the information flow, while at the same time retaining an amount of structural entropy such that the network can encode and express sufficient diversity through its traits. 

Thus, to complete the analogy to thermodynamics, the trade-off between the diversity of information and information flow is captured by the efficiency~\citep{Ghavasieh2024}:

\begin{equation}
    \label{eq:efficiency}
    \eta = \frac{W + Q}{W} = 1- \frac{|Q|}{W} = \frac{U}{W}
\end{equation}
where $0 \leq \eta \leq 1$ and $U \equiv - \partial_\tau \log Z \equiv \Tr [\hat{L} \hat{\rho}]$.
The variational principle here exposed has been devised to macroscopically explain the sparsity and its empirical scaling law, as well as the emergence of topological features such as modularity, small-worldness, and, with partially less clear results, heterogeneity~\citep{Ghavasieh2024}.
In our work, we use this variational principle to explain the structural properties of the network subclass composed of heterogeneous hierarchical trees. The specific heat is used to identify the resolution scale at which such a variational principle is effective for our subclass of networks during the transition to the scale-invariant regime. For scale-invariant networks, for some $\tau$-scales, $U=<\lambda>_\tau = C_0 / \tau$~\cite{Villegas2023,poggialini2024networks}, hence in this regime \Cref{eq:efficiency} becomes $\eta = \frac{1}{\tau}\frac{C_0}{W}=\frac{1}{\tau}\frac{d_s}{2 W}$. An additional element of this work is that the investigated trees are required to encode the diversity expressed in the environment they are in contact with, defined by a real-world dataset of patterns, $\mathcal{D}$.

\subsection{Portegys algorithm}
\label{sec:portegys_alg}

\subsubsection{Network growth process}
\label{sec:network_growth}

To generate structures based on real-world disorder, we use the algorithm introduced by Portegys~\citep{Portegys} for accelerating nearest-neighbor (1-NN) pattern recognition tasks and subsequently extended to k-NN~\cite{Mangalagiu1999}. The Portegys algorithm we use is the extended version with $k=5$, where the distance between patterns is the Euclidean distance. 

Given the external constraints, the single control parameter, $r$, and a high-dimensional vector collection, the dataset $\mathcal{D}$, the system seeks the best-fitness configuration to map the dataset's disorder into a tree, i.e. a network that is loop-less and sparse, $|E| = N-1$, where $|E|$ is the number of links, and $N$ the number of nodes.

The tree-formation protocol reorganizes raw data using a recursive method in which samples are randomly selected and inserted into the growing tree network.
At each step, the new pattern is linked to either the current parent pattern or to one of the k-NN children of the parent pattern. The choice is based on the distance between the parent and the new pattern and the one evaluated between the child pattern and the new pattern. The parameter $r$ modulates the relative importance of these two distances creating a larger or smaller region of influence of the parent in the sample space.



\begin{figure}[ht!]
\begin{center}
\includegraphics[width=1\columnwidth]{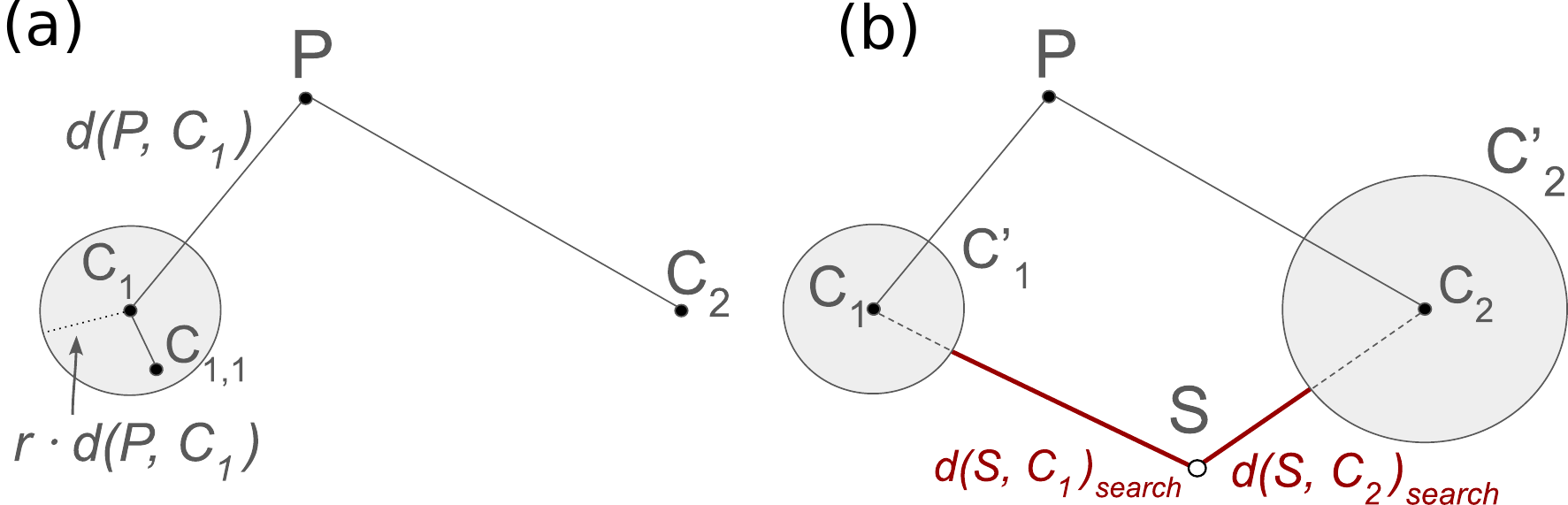}
\end{center}
\caption{\textbf{Search method based on relative distance}. In \textbf{(a)} the growth process is illustrated. At each step, a new pattern is linked to either the current parent pattern, $P$, or to one of its children, $C_i$. The parameter $r$ modulates the region of attraction of the child, indicated by the gray circle around $C_1$, and the one of the parent node. Points falling within the distance $r\cdot d(P, C_i)$ are linked to the child, otherwise, an edge will be defined between the parent node, $P$, and the new pattern. In \textbf{(b)} the search method based on relative distance is illustrated. Once a search pattern arrives at node $P$ it must choose to search either the subtree of $C_1$ or $C_2$. The choice is based on minimizing the distance to the expected patterns contained in the subtree of $C_i$. In the depicted case, patterns closer to the search pattern, $S$, are expected to be in the subtree of $C_2$. Adapted from \cite{Schomaker2000-dt}.}
\label{fig:growth_search_alg}
\end{figure}

\subsubsection{Search phase}
\label{sec:search_phase}

After the tree formation described in~\Cref{sec:network_growth}, the Portegys algorithm enters the search phase, where the best match to an unknown search pattern, $S$, is given using a first-best approach. The original goal of this algorithm was to restructure a list of examples in a tree-based way, such that the matching accuracy would not be much worse than nearest-neighbor matching on a full list while providing a considerable speed up due to the hierarchical organization into a tree-based structure. The growth process, described in \Cref{sec:network_growth} and depicted in \Cref{fig:growth_search_alg}(a), ensures that subtrees of a given node have on average decreasing distances for decreasing hierarchical levels. 

This provides the possibility of using a search technique based on relative distances. The search method is depicted in \Cref{fig:growth_search_alg}(b) and is an iterative process that begins at the root-node pattern. Once a search pattern, $S$,  arrives at a certain parent node, $P$, it must decide to continue to search for the best match into one of its children subtrees, $C^\prime_1$, $C^\prime_2$ in \Cref{fig:growth_search_alg}(b). The choice is taken minimizing the Portegys search-distance: $d(S,C_i)_{search}= d(S, C_1) - r \cdot d(P,C_i) = d(S,C^\prime_i)$, which quantifies the distance between the $S$-pattern and the potential patterns appearing in the subtree of each child, $C_i$, collectively indicated by the gray circle, $C^\prime_i$, in \Cref{fig:growth_search_alg}(b). Where $d(\cdot)$ indicates any metric distance equipped with semi-positivity, symmetry, and triangular inequality properties. If $d(S,C_i)_{search}$ is less than zero then it is set to zero. In \Cref{fig:growth_search_alg}(b) the bold segments in red indicate the search distance to each child node and equivalently the expected distance between the unknown pattern $S$ and their expected neighborhoods of patterns $C^\prime_i$.

\subsection{Datasets}
\label{sec:datasets}
Our results are evaluated over subsets of four different datasets ranked based on their complexity (\Cref{sec:dataset_complexity}): the NIST~\cite{NISTSD}, the MNIST~\cite{mnist}, the FashionMNIST~\cite{fashionMnist}, and the CIFAR10~\cite{cifar10} datasets. These datasets contain randomly collected instances of image samples out of 10 classes. It should be noted that in this type of data, a pattern class can generally not be represented by a single centroid vector, but will contain multiple densities (``sub-style patterns'') such that the sample point cloud can be ordered naturally in a network structure.

The MNIST database contains $24 \times 24$ grey-scale images of handwritten digits [0-9] commonly used to benchmark machine- and deep-learning algorithms.  The NIST dataset is the ancestor of the MNIST, it contains $16 \times 16$ black-and-white images of the handwritten digits [0-9]. The FashionMNIST is a large dataset of $28 \times 28$ gray-scale images of fashion products from 10 categories. The CIFAR-10 dataset contains $32\times32$ color images belonging to 10 classes. In all our experiments a subset of 2000 samples is extracted from the train set and the test set independently. Samples from the FashionMNIST and MNIST are always binarized. The samples from the CIFAR10 are normalized between 0 and 1 dividing each pixel value by 255, and values are rounded up to eighth decimal. In the Supplemental Material~\cite{SM}, we show results for larger networks generated from $4000$ samples from the MNIST and FashionMNIST. The 10 classes are equally represented in both the test and training set.

\section{Results}
\label{sec:results}

\subsection{Dataset complexity}
\label{sec:dataset_complexity}

\begin{figure}[hbtp!]
\begin{center}
\includegraphics[width=1\columnwidth]{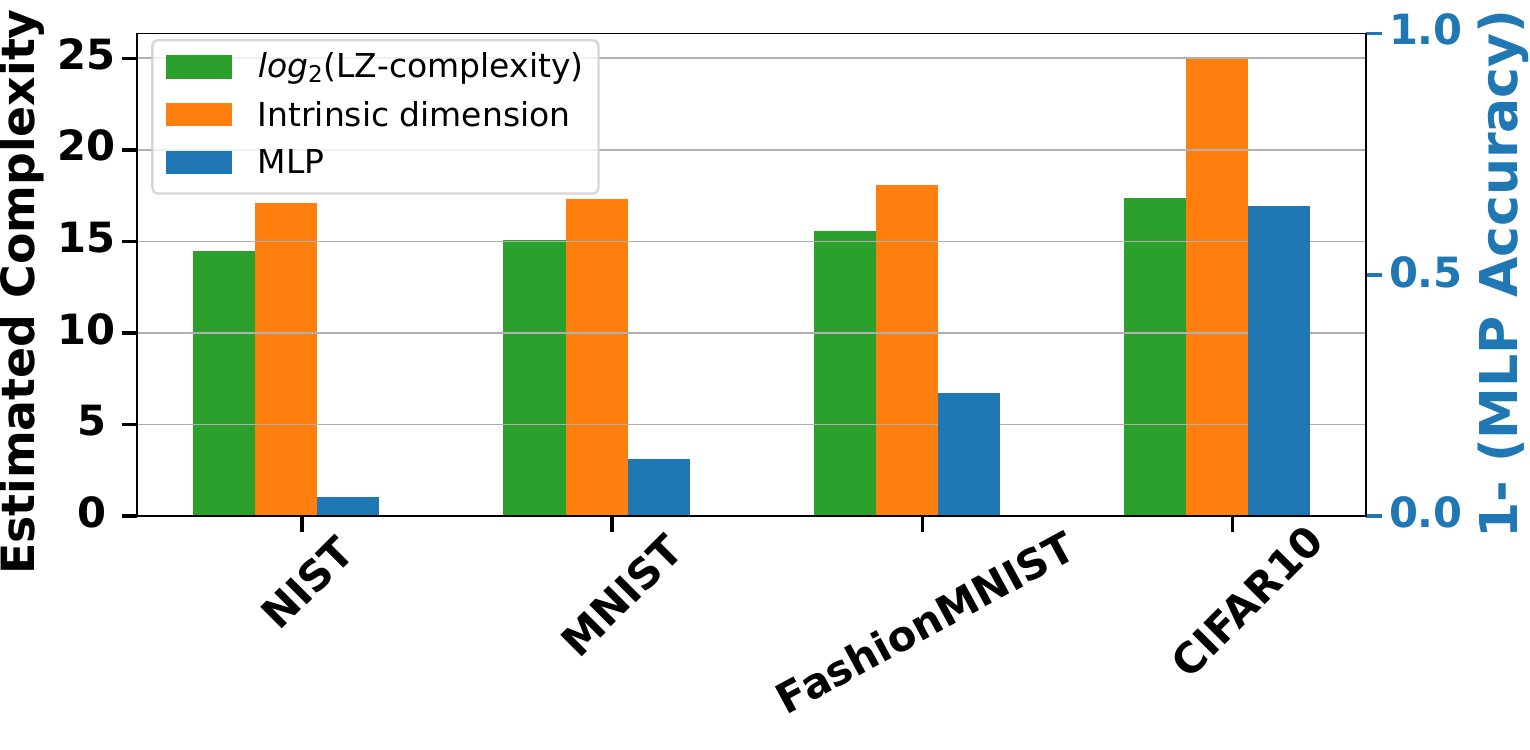}
\end{center}
\caption{\textbf{Dataset complexity} is estimated with multiple methods for each training dataset of 2000 samples encountered in this work. In green an estimate is given computing the Lempel-Ziv complexity~\cite{Lempel1976OnTC}, while in orange the intrinsic dimension is estimated with the Maximum Likelihood Estimation approach~\cite{levina2004maximum, Mackay_Ghahrami,pope2021intrinsic}. In blue the accuracy on the test set for a multilayer-perceptron (MLP) with two hidden layers of 128 units each. The left axis refers to the quantities in green and orange, while the right axis refers to the multilayer-perceptron accuracy. The CIFAR10 dataset is estimated as the most complex, while the NIST dataset is the least complex among the four.}
\label{fig:dataset_complexity}
\end{figure}

The datasets used in our experiments can be ranked based on their estimated complexity, see \Cref{fig:dataset_complexity}. We estimate each dataset's complexity with multiple methods on the same subset of $2000$ training samples used in the rest of the manuscript. 

The first heuristic approach is based on the test accuracy of a small multilayer perception (MLP) (2 layers with 128 units each). In pattern recognition, the complexity of a dataset is determined by the complexity of the curvilinear hyperplanes that would be needed to optimally separate patterns from different classes. A simple but effective parameter to gauge this facet of complexity is the accuracy that a classifier (such as the MLP) can achieve in determining the true class label for a data point. Thus, a more complex dataset would result in a lower accuracy for neural networks of fixed size when compared to a less complex dataset. 

The second method we use is the computation of the Lempel-Ziv complexity~\cite{Lempel1976OnTC} of the aggregated training set. In this case, the samples need to be binarized. 

The third method concerns the measure of the intrinsic dimension~\cite{mezard2023spin,pope2021intrinsic} of the dataset using the Maximum Likelihood Estimation approach introduced in Ref.~\cite{levina2004maximum} with the correction proposed in Ref.~\cite{Mackay_Ghahrami}. A given dataset of gray-scale images of size $pixel \times pixel$ a priori lies on a high-dimensional space of $\mathbb{R}^{pix \times pix}$, nevertheless, because it might contain a specific class of objects that account for the ``world'' of our problem, for instance, handwritten digits, it spans a lower dimensional manifold. Even though the definition of such a manifold is an open problem, it is possible to estimate its dimension based on the distance of the $n$-neighbouring points~\cite{mezard2023spin,pope2021intrinsic}. We set the hyper-parameter $n=3$.

\subsection{Pattern-matching efficiency}
\label{sec:pattern_matching_efficiency}

Mapping a high-dimensional vector collection into a hierarchical tree structure limits the number of comparisons needed to match an unknown pattern to a prototypical example. Namely, the algorithm introduced by Portegys~\cite{Portegys,Mangalagiu1999} is equipped with a pattern-searching method illustrated in \Cref{fig:growth_search_alg} and described in~\Cref{sec:search_phase}, representing the functional aspect of the system.

\begin{figure}[ht!]
\begin{center}
\includegraphics[width=1\columnwidth]{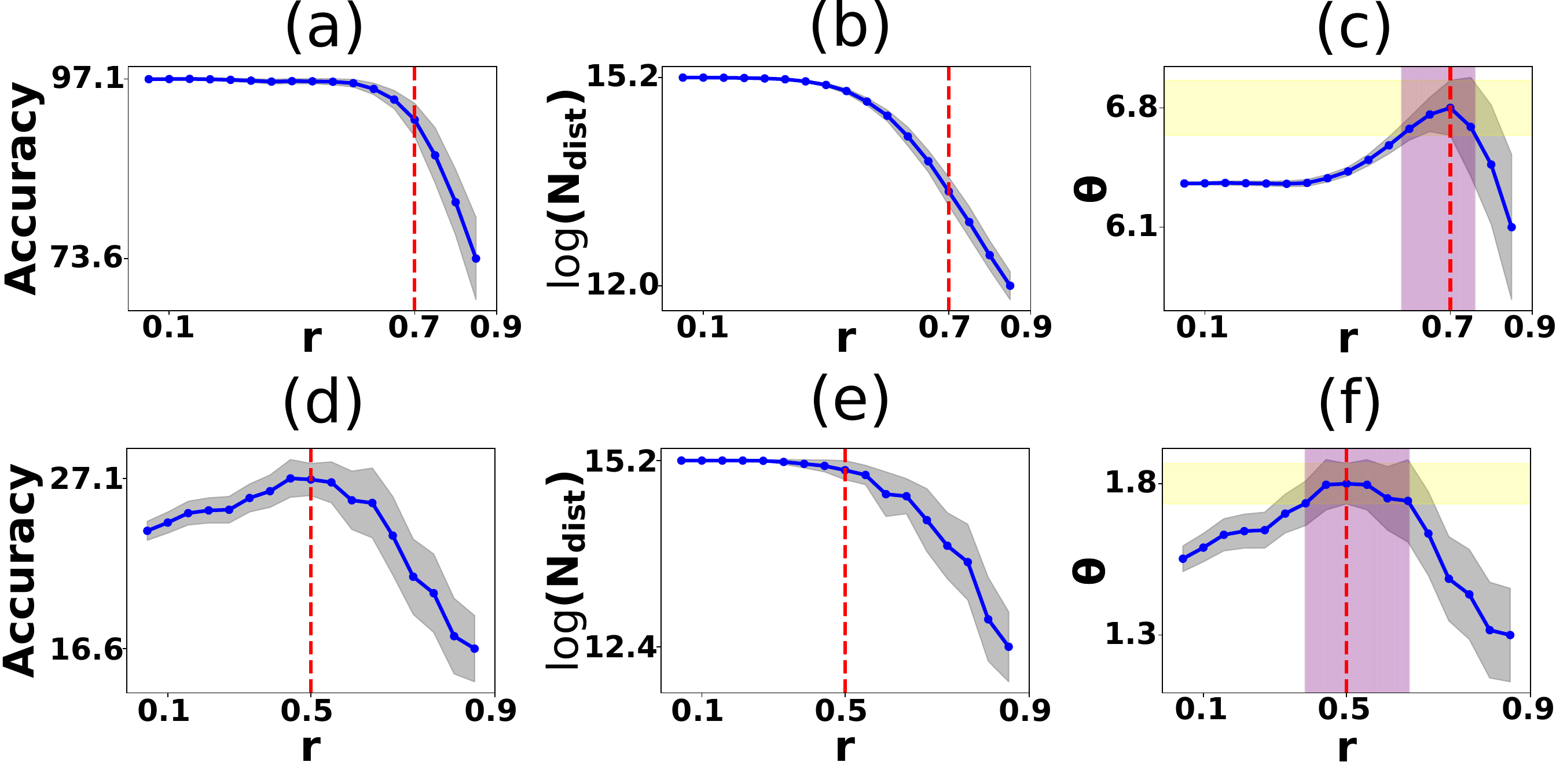}
\end{center}
\caption{\textbf{Pattern-matching efficiency} in \textbf{(c),(f)} measured over the NIST and CIFAR10 respectively. The efficiency, $\theta$, is defined as the trade-off between the classification accuracy, \textbf{(a),(d)}, and the logarithm of the number of distances computed to find the best match, \textbf{(b),(e)}. The dashed-red vertical line indicates the optimal $r$ value in terms of efficiency, $\theta$. The shaded grey area designates the standard deviation. The purple shaded area in panel \textbf{(c),(f)} indicates a tolerance $r$-domain of optimal efficiency. The tolerance domain is defined as the $r$-values such that the corresponding $\theta$-values are included in one standard deviation computed at the maximum $\theta$, whose y-axis projection is indicated by the yellow shaded area. Curves are computed over 50 independent simulations with a subset of 2000 samples extracted from the NIST and CIFAR10 datasets. The 10-digit classes are equally represented both in the train- and test sets.}
\label{fig:pattern_matching_efficiency}
\end{figure}

We define a pattern-matching efficiency, $\theta$, in terms of class match accuracy, $A$, and the number of distances computed, $N_{dist}$:

\begin{equation}
\label{eq:theta}
    \theta = \frac{\text{A}}{\log N_{dist}}
\end{equation}
the logarithm is taken since the range of $N_{dist}$ can virtually span the interval $N_{dist} \in (0, N]$, where $N$ is the number of nodes. For the case of a star graph, $r \rightarrow 0$, the number of computed distances is maximum $N_{dist}=N$; the number of computed distances decreases for increasing $r$, potentially lowering the class-match accuracy. 
Nevertheless, we highlight \Cref{fig:pattern_matching_efficiency}(d) and note that the CIFAR10 dataset exhibits the global accuracy peak at an intermediate value of $r$. We interpret this observation by saying that for some datasets (see also the case of the FashionMNIST in Sec. S7 of the Supplemental Material~\cite{SM}) specific network structures do possess the ability to better encode information of the ``world'' in their traits. We postpone the discussion of this finding to \Cref{sec:results,sec:discussion}.

When defining the pattern-matching efficiency, we are making the simplified assumption that the pattern-matching algorithm runs on a physical substrate such that visiting any node has a fixed cost in terms of energy or in terms of general resources. Hence, the efficiency, $\theta$, quantifies the trade-off between the cost of used resources and the gain in performance, as depicted in~\Cref{fig:pattern_matching_efficiency}(c),(f) for the MNIST and CIFAR10 respectively. 

While in \Cref{sec:phasediagram} we discuss the full range of $r$. In the rest of the manuscript, we limit the discussion to $r\leq0.85$ as the accuracy, and hence the pattern-matching efficiency, quickly drops for larger $r$, moreover, the algorithm shows instability as it doesn't necessarily converge for $r \geq 0.9$ and $N=2000$.

\subsection{Phase portrait}
\label{sec:phasediagram}

\begin{figure*}[ht!]
\begin{center}
\includegraphics[width=\textwidth]{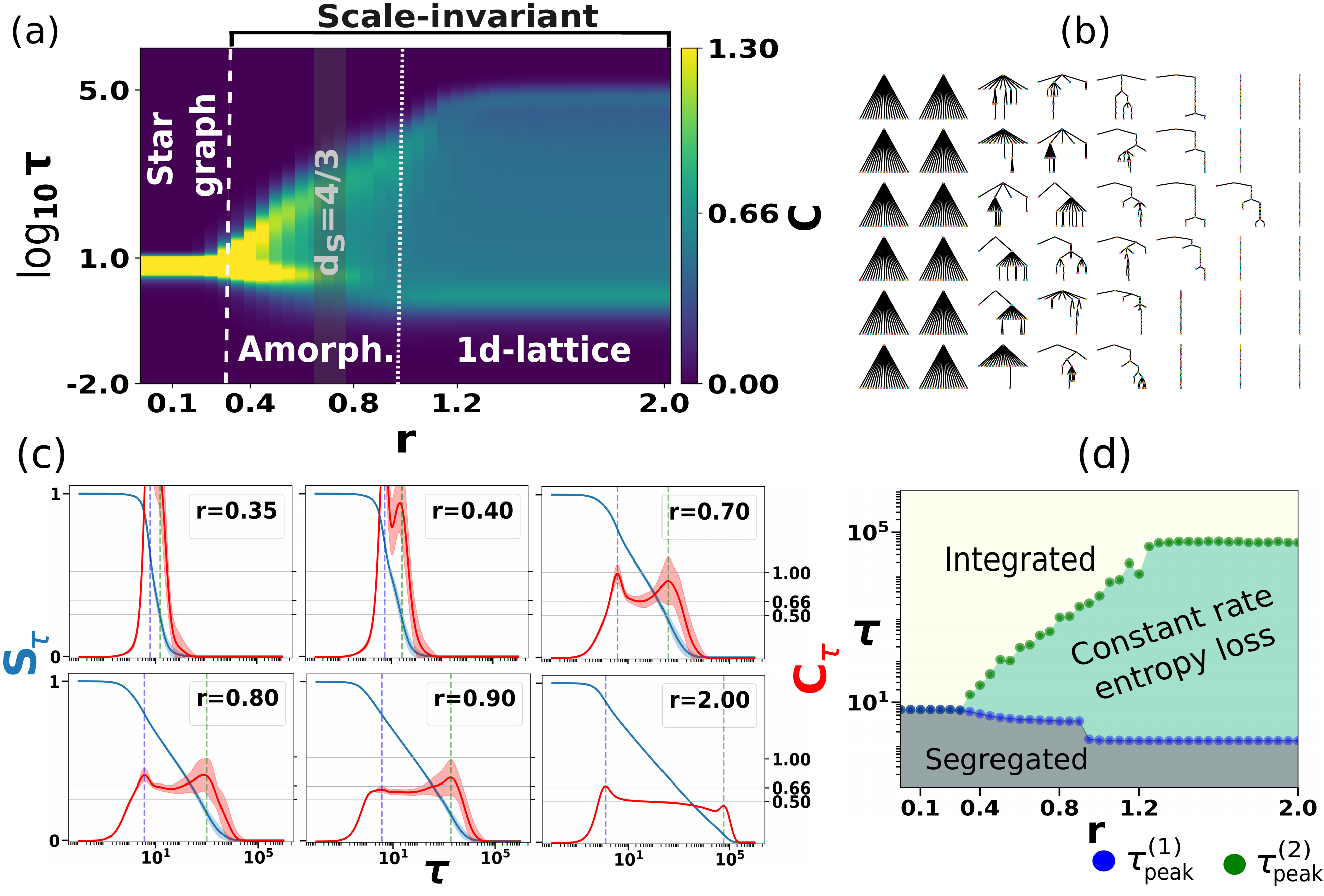}
\end{center}
\caption{\textbf{Multiple phase transitions in hierarchical tree networks.} In \textbf{(a)} the specific heat, $C$, across the range $r \in [0,2]$. At $r \simeq 0.35$, indicated by the dashed white line, the hierarchical network transits from the star graph to the amorphous configuration with the emergence of scale-invariant traits evidenced by a plateau in the specific heat. The scale-invariant region shows a second phase transition from amorphous to regular 1d-lattice at $r\simeq1.05$ indicated by the dotted white line. For $r>0.35$, the network is always scale-invariant with decreasing spectral dimension up to the transition point, above which the spectral dimension saturates to $d_s = 1 = 2 \cdot C_0$, where $C_0$ is the plateau height in (c). Note that at $r\simeq 0.7$ the spectral dimension is $d_s=4/3$ as expected for random trees with finite variance in the node degree. In panel \textbf{(c)} the entropy (blue) and the specific heat (red) for $r=\{0.35,~0.40,~0.70,~0.80,~0.90,2.00\}$. Vertical blue and green lines indicate the $\tau$-scale of the first and the second peak, $\tau^{(1,2)}_{peak}$, of the specific heat respectively. In \textbf{(d)}, the $\tau$-scales of the two specific heat peaks are illustrated with blue and green dots respectively delimiting the three dynamical phases: the segregated phase (grey), the integrated phase (light yellow) and the green region where \textit{self-similar structures} are integrated at a constant-rate across the network resolution scales. In \textbf{(b)} exemplary hierarchical trees of $N=20$ for growing $r$ from left to right. All curves in this figure are realized for networks of $N=500$ randomly sampled from the FashionMNIST dataset and averaged over 50 independent realizations.}
\label{fig:phase_diagram}
\end{figure*}

\begin{figure*}[ht!]
\begin{center}
\includegraphics[width=.85\textwidth]{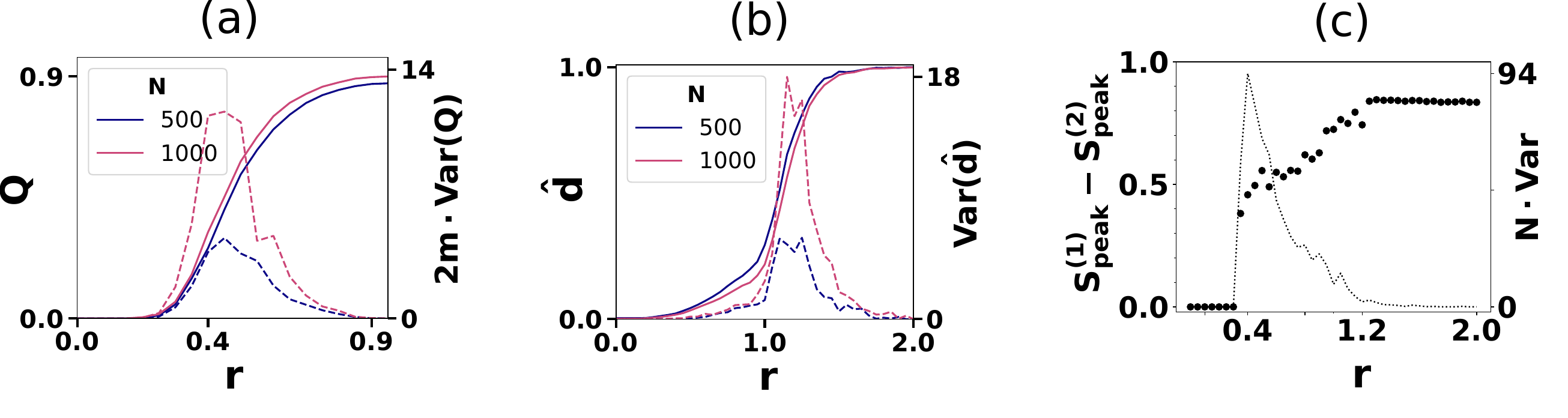}
\end{center}
\caption{\textbf{Order parameters.} In \textbf{(a)}, the modularity, $Q$, resolves the transition involving the mesoscale structure. In \textbf{(b)}, the normalized diameter, $\hat{d}=d/m$, resolves the large-scale transition. Their respective variances (dashed lines) identify the transition points. In \textbf{(c)}, the difference of entropy at the two $\tau^{(1,2)}_{peak}$ peaks quantifies the normalized (log)number of informationally connected nodes across the scale-invariant $\tau$-domain, i.e. the number of nodes included in the scale-invariant region. Note the discontinuity at the transition from star to amorphous. 
Curves in (a) and (b) are realized for networks of $N=500,~1000$ randomly sampled from the FashionMNIST dataset and averaged over 50 independent realizations. Curves in (c) are for networks of $N=500$.}
\label{fig:order_parameter}
\end{figure*}

The Portegys algorithm produces three prototypical structures as those depicted in \Cref{fig:structures} and \Cref{fig:phase_diagram}(b), depending on the value of the control parameter, $r$ (see in Methods \Cref{sec:portegys_alg}). 

For $r \rightarrow 0$, every node is linked to the root node and a star graph turns out (see in \Cref{fig:structures} the central node, which is the first node selected in the growth process). For some intermediate value of $r$, a ragged tree structure with progressively smaller distances between the patterns is produced; while for $r \rightarrow 2$, a 1d-lattice is created.

Two structural phase transitions involving two different resolution scales can be identified (\Cref{fig:phase_diagram}(a)). The transition at the largest resolution scale can be described using as an order parameter a quantity that describes the large-scale network properties: $\hat{d}= d/m$, where $d$ is the diameter of the tree and $m$ is the number of edges equal to $m=N-1$. For hierarchical trees, another choice would be the normalized number of hierarchical levels $\hat{h} = h/(N-1)$ where $h$ is the number of hierarchical levels ranging from $h \in [0, N-1]$ (see illustration in \Cref{fig:order_parameter}(b)). Both quantities describe the ``volume'' expansion controlled by the parameter $r$. We chose the normalized diameter $\hat{d}$ and we identify this phase transition at $r \simeq 1.05$ as indicated by the peak of the variance (dotted lines) in \Cref{fig:order_parameter}(b) for the two network sizes $N=500,~1000$. This phase transition border distinguishes between networks in the amorphous regime from those with regular 1d-lattice structure. Note that on the left-hand side of the transition point, at $r\simeq 0.7$ the spectral dimension is $d_s=4/3=2C_0$, typical of random trees with finite variance in the node degree's distribution~\cite{Destri_2002}, where $C_0$ is the specific heat plateau height (see also \Cref{fig:phase_diagram}(a),(c) and Fig. S1 in the Supplemental Material~\cite{SM}). For $r\gtrsim 1$, $d_s=1$, as it is expected for a 1d-lattice.

The second change of phase is the transition from the star graph to the amorphous configuration at $r\simeq 0.35$. Here, the emergence of scale-invariant structural properties is reflected by the progressively larger specific heat plateau in between the two $C$-peaks, as shown in \Cref{fig:phase_diagram}(a),(c). For $r \rightarrow 2$, the plateau would include the full 1d-lattice in the thermodynamic limit $N \rightarrow \infty$. 

Before the transition, a single peak in the specific heat identifies a single trivial $\tau$-resolution scale below which the informationally connected components are the single nodes, and after which the informationally connected component is the full graph. After the phase transition, a second peak of the specific heat emerges at larger $\tau$ scales. This second peak of $C$ in \Cref{fig:phase_diagram}(c) identifies the scale at which the information is integrated between the distinct modules. In other words, the second peak reflects the ``degree'' of the large-scale modules, in much analogy to the first peak, which reflects the presence of hubs at the local connectivity scale of the nodes. Note that we exclude both the clusters, i.e. triangles, and higher-order cycles from our discussion, as the tree networks we address are purely loop-less.

The star graph and amorphous network phase space regions are those investigated in this work. Because the peaks in the thermodynamic and pattern-matching efficiencies always occur in the amorphous and scale-invariant region of the phase plane. In some cases, they appear in the vicinity of the transition border between the two at $r\simeq 0.35$. This transition, conversely to what we did with the amorphous to regular-lattice transition discussed above, can’t be identified by a quantity resolving the large-scale property of the network. Indeed, we observe in \Cref{fig:order_parameter}(b) the lack of a peak in the variance of $\hat{d}$, which would indicate no sign of \textit{bona fide} phase transition. Nevertheless, a phase transition exists and an appropriate order parameter needs to be employed to identify it.

This phase change occurs at smaller resolution scales with respect to the first phase transition we discussed. Specifically, it involves the mesoscales properties changing from the star graph to the scale-invariant regime (see the bifurcation of the specific heat in \Cref{fig:phase_diagram}(a),(c)). Thus, we can identify this transition using as an order parameter the modularity~\cite{Newman2006}, $Q$, of the tree which we show in \Cref{fig:order_parameter}(a). We are taking advantage of the fact that the modularity for trees is bounded between $[0, 1]$ and that in the case of a 1d-lattice, the modularity is equal to $1$ if we partition the network into $N$ communities, which is the case if we use the community detection algorithm introduced in Ref.~\cite{Blondel2008}. We stress that modularity is a mesoscale descriptor while the diameter of the network is a large-scale descriptor. Thus, a mesoscale resolution is required to monitor the structural changes occurring across the star graph phase to the amorphous scale-invariant phase transition. This motivates the choice of the mesoscale $\tau \sim 10$ employed in \Cref{sec:results}.

As a second possible order parameter, for the mesoscale phase-transition, we show in \Cref{fig:order_parameter}(c) the difference of entropy at the two $\tau^{(1,2)}_{peak}$ specific heat peaks. This order parameter counts the normalized (log)number of informationally connected nodes across the $\tau$-scales of the specific heat plateau. In simple terms, we count the fraction of nodes included in the scale-invariant region. We observe a discontinuity at the border of the scale-invariant transition that is not observed at the border of the amorphous to regular lattice transition.

As already discussed in \Cref{sec:statistical_phys}, also along the $\tau$-direction an informational phase transition exists~\cite{Villegas_2022,Villegas2023} delimited by the peaks of $C$ at $\tau^{(1,2)}_{peak}$. Using the specific heat peaks, we can distinguish three \textit{dynamical} phases (see \Cref{fig:phase_diagram}(d)). For $\tau < \tau^{(1)}_{peak}$, the network is in the informationally segregated phase; while for $\tau > \tau^{(2)}_{peak}$ is in the informationally integrated phase. Only for $r > 0.35$ and $\tau^{(1)}_{peak} < \tau < \tau^{(2)}_{peak}$, we observe the \textit{dynamical} regime of \textit{constant-rate entropy loss} characteristic of scale-invariant structures with dilation symmetry~\cite{Villegas2023,poggialini2024networks}.

\begin{figure*}[ht!]
\begin{center}
\includegraphics[width=\textwidth]{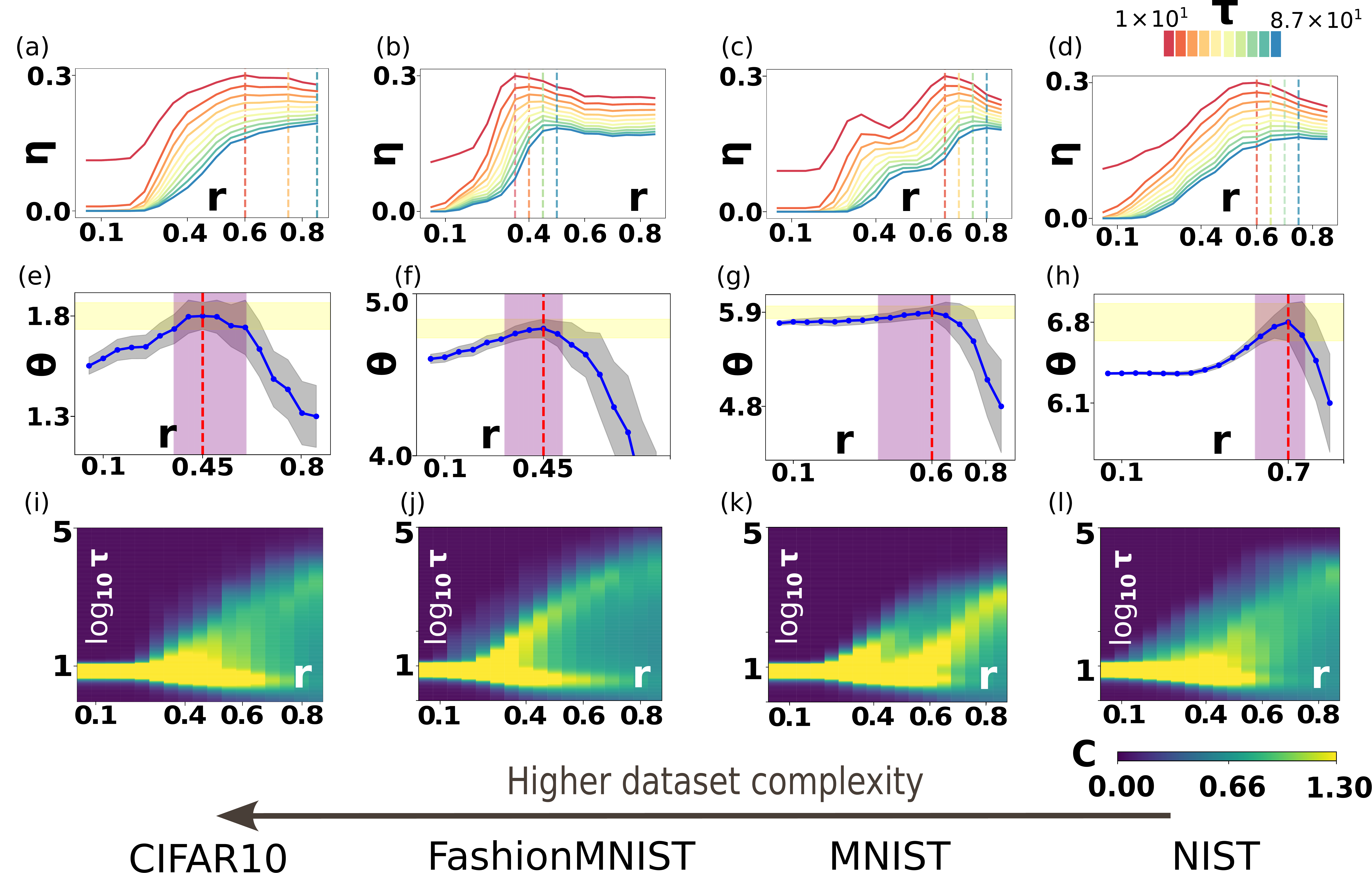}
\end{center}
\caption{
\textbf{Thermodynamic \& pattern-matching efficiency}. 
In \textbf{(a--d)} the thermodynamic efficiency, $\eta$, as a function of $r$. The thermodynamic efficiency is measured in the $\tau$-range of one order of magnitude from the bifurcation of the specific heat, $C$, depicted in panels \textbf{(i--l)}. The $\tau$-resolution scale $\tau \sim 10$ identifies the relevant ``length'' scale resolving the transition from the single-peak to the emergence of a roughly plateau region, $C=C_0>0$, reflecting a roughly scale-invariant structure.
In \textbf{(e--h)} the pattern-matching efficiency, $\theta$, is plotted as a function of $r$ for the four investigated datasets of increasing complexity (see arrow at the bottom of the figure). The shaded purple area indicates the tolerance $r$-range for the optimal pattern-matching efficiency, which includes points in one standard deviation from the global maximum indicated by the vertical red-dashed line. In panels (b),(f) and (d),(h) a noticeable cohesion of the peaks of the two efficiencies is observed. Vertical dashed lines indicate the global maxima of the efficiencies.
The efficiency peaks, both $\eta$ and $\theta$, are always in the amorphous region of the phase plane. For datasets of higher complexity, they tend to be closer to the structural phase transition border separating the star graph from the amorphous region. 
This agrees with the argument that a fairly simple dataset does not benefit from the vicinity of the phase transition. Quantities are averaged over 50 independent simulations.}
\label{fig:table_FMINIST_MNIST_NIST_2000}
\end{figure*}

\subsection{Efficiency}
In \Cref{fig:table_FMINIST_MNIST_NIST_2000}, panels (a--d) depict the thermodynamic efficiency, $\eta$, and panels (e--h) depict the pattern-matching efficiency, $\theta$. See \Cref{eq:efficiency,eq:theta} respectively.

The thermodynamic efficiency~\cite{Ghavasieh2024}, $\eta=(Q+W)/W$, quantifies the trade-off between two quantities. 
The former, $Q=-\delta S/\tau$, is proportional to the variation of entropy content, $\delta S$, from the case of isolated nodes (see Methods). Thus, it estimates the ability to express diversity to a stochastic perturbation, or in other terms, the internal representation capabilities provided by the network topological traits. 
The latter is the variation of information flow with respect to a network of isolated nodes, $W=\delta F$; so to speak, how fast information travels across the network (see Methods).  

The pattern-matching efficiency, $\theta$, defined in \Cref{eq:theta}, quantifies the trade-off between accuracy and number of visited nodes in the pattern-recognition task, representing the functional aspect of the algorithm (see Methods).

The vertical dashed lines indicate the peaks of the efficiencies, while the shaded purple regions in panels (e--h) indicate a $r$-domain tolerance for $\theta$. The tolerance domain is defined as those $r$-values such that the corresponding $\theta$-values are included in one standard deviation computed at the best trade-off. 

The two efficiencies $\theta$ and $\eta$ show some consistency in the x-range of highly efficient structures, see panels (b),(d) related to the NIST and the FashionMNIST. If we include points in the tolerance $r$-range spanned by the purple-shaded areas in panels (e--h), we observe the same consistency for all datasets when $\eta$ is computed at $\tau=10$ (red dashed lines in panels (a--d)).
This consistency in the maxima suggests that the two efficiencies might be related, in the specific context of the algorithm here investigated. In the \Cref{sec:discussion} we discuss this matter. 

\begin{figure*}[ht!]
\begin{center}
\includegraphics[width=\textwidth]{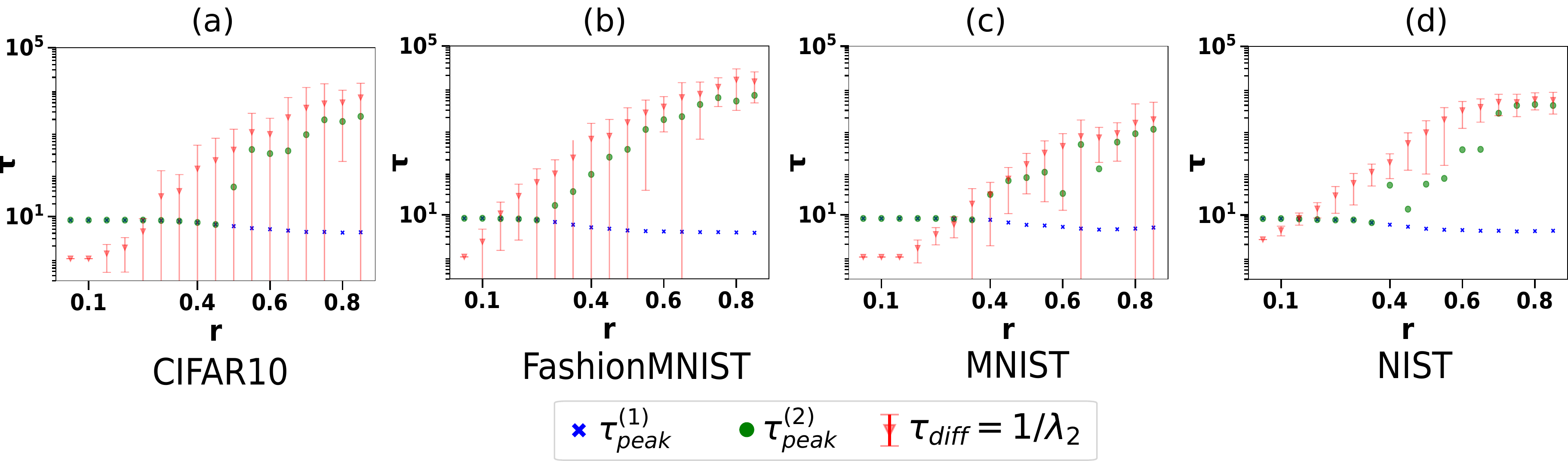}
\end{center}
\caption{\textbf{Role of the ensemble average on the estimation of the $\tau$-scale}. Panels \textbf{(a--d)} depict the $\tau^{(1,2)}_{peak}$
and $\tau_{diff}$ as a function of $r$. $\tau^{(1,2)}_{peak}$ are the $\tau$-location of the first and last peak of the specific heat in \Cref{fig:table_FMINIST_MNIST_NIST_2000}. $\tau_{diff}$ is the inverse of the Fiedler eigenvalue which provides a good estimation of the end of the diffusion process only for sufficiently large $r$. Note that the red curves retrace the profile of the upper branch of the specific heat for $\tau \gtrsim 10$ in panels (i--l) of \Cref{fig:table_FMINIST_MNIST_NIST_2000}. At smaller $r$ values the equilibrium time is better reflected by $\tau^{(2)}_{peak}$. The equilibrium of the diffusion process is an increasing function of $r$. $\tau_{diff}$ curves are averaged over 50 independent simulations. The $\tau^{(1,2)}_{peak}$ curves are computed from the specific heat curves which are averaged over 50 independent simulations.}
\label{fig:tau_FMINIST_MNIST_NIST_2000}
\end{figure*}

The specific resolution scale, $\tau \sim 10$, at which we compute $\eta$ is the scale involved in the mesoscale structural transition from star graph to amorphous scale-invariant, that we discussed in~\Cref{sec:phasediagram} and \Cref{fig:order_parameter}. Such critical resolution scale is identified by the bifurcation of the specific heat peaks in \Cref{fig:table_FMINIST_MNIST_NIST_2000}(i--l) and in \Cref{fig:tau_FMINIST_MNIST_NIST_2000}(a--d).

In the seminal work in Ref.~\citep{Ghavasieh2024}, it is suggested to use as a reference the $\tau$-scale equal to the inverse of the Fiedler eigenvalue, i.e. the smallest non-zero Laplacian eigenvalue $\lambda_2 = 1/\tau_{diff}$, and logarithmically divide $\tau_{diff}$ to define the large propagation scales, the middle scales, and the small scales where to compute the thermodynamic efficiency, $\eta$. 
The Fiedler eigenvalue provides the largest network time scale where the diffusion process has reached equilibrium.
Nevertheless, $\tau_{diff}$ seems to only approximate the end of the diffusion process which is better reflected by $\tau^{(2)}_{peak}$, especially in the star-graph region for small $r$. Panels (a--d) of \Cref{fig:tau_FMINIST_MNIST_NIST_2000} depict the $\tau_{diff}$ curves in red, next to the $\tau^{(1,2)}_{peaks}$ curves. The key difference is the thermal ensemble average: a weighted average, that employs all the eigenvalues to compute the specific heat and identify its peaks.

Differently from what was proposed in Ref.~\cite{Ghavasieh2024}, we don't use $\tau_{diff}$, or $\tau_{peak}^{(2)}$ in \Cref{fig:tau_FMINIST_MNIST_NIST_2000}. Because that would mean adopting the largest possible resolution, which would not monitor the structural mesoscale changes involved in the transition, from the single trivial scale of the star graph to the more complex topological traits of the amorphous networks in the scale-invariant regime.

The resolution scale, $\tau\sim10$, is shared across all topologies in the phase plane and, remarkably, defines an ``ultra-violet'' cut-off in the characteristic frequency of the networks across the transformation from one phase to the other. This is in analogy to the process of integrating out the large eigenvalues responsible for the finer details, as discussed in the context of the Laplacian-Renormalization group~\cite{Villegas2023}. Note in Fig. S1 of the Supplemental Material~\cite{SM} that our choice of $\tau=1/\lambda\sim10$ is supported by the exclusion of the resolution region where the power-law behavior of the spectral density, $\mathcal{P}(\lambda)$, is lost. In Sec. S3 of the Supplemental Material~\cite{SM}, the efficiency, $\eta$, is computed in the range interval $\tau < 10$. Whether computed at those scales, $\eta$ shows a change in concavity and no relation with the pattern-matching efficiency.

Moreover, we observe that, while the value of $r$ where a structural transition appears is dependent on the dataset, i.e. $r=0.40$ for the NIST, $r\simeq 0.40$ for the MNIST, $r \simeq 0.35$ for the FashionMNIST, and $r \simeq 0.50$ for the CIFAR10, the $\tau \sim 10^1$ scale is the same for all the investigated datasets. 
Finally, we note that the bifurcation $\tau$-scale is also unaffected by the number of nodes in the network (see Supplemental Material~\cite{SM} Sec. S2).

\subsection{Datasets' phase plane}
The topological complexity of the networks increases at a critical $r$, from exhibiting the trivial traits of a star graph to the structure of the amorphous network with a roughly power-law spectrum. Thus, in the phase plane, the emergence of structure is reflected by the rise of a second peak in $C$ together with a roughly $C=C_0>0$ plateau between the first and the last peak. This is illustrated in \Cref{fig:table_FMINIST_MNIST_NIST_2000}(i--l) for each investigated dataset.
Such transition is also depicted in \Cref{fig:tau_FMINIST_MNIST_NIST_2000} by the bifurcation of the blue and green curves for $\tau^{(1)}_{peak}$ and $\tau^{(2)}_{peak}$ indicating the lower and upper branches in \Cref{fig:table_FMINIST_MNIST_NIST_2000}(i--l).
See also Sec. S4 in the Supplemental Material~\cite{SM} which illustrates the spectral entropy and specific heat curves for the four datasets independently depicted for each $r$-value.

The MNIST, an intermediate dataset in terms of complexity, is an interesting case. We note that the specific heat of the MNIST has a third large peak in between $\tau^{(1)}$ and $\tau^{(2)}$ (see \Cref{fig:table_FMINIST_MNIST_NIST_2000}(k)). This is an intrinsic characteristic of the dataset, as this feature is only mitigated for larger networks, as shown in Figs. S2 and S7 of the Supplemental Material~\cite{SM}.

This intermediate specific heat peak is reflected on two maxima for the thermodynamic efficiency, $\eta$, and a two-step transition in the amorphous region, in \Cref{fig:table_FMINIST_MNIST_NIST_2000}(c),(k). 
The local maximum is at $r\in [0.35-0.45]$, depending on the specific $\tau$-scale. Such a range aligns with the transition at $r \simeq 0.40$ indicated by the bifurcation of the two specific heat peaks in \Cref{fig:tau_FMINIST_MNIST_NIST_2000}. 
The global maximum in \Cref{fig:table_FMINIST_MNIST_NIST_2000}(c) is close to the second-step transition at $r\simeq0.6$. 
Although the local maximum in $\eta$ is not detected by $\theta$ in \Cref{fig:table_FMINIST_MNIST_NIST_2000}(g), it is reflected in its standard deviation, allowing us to include the $r$-range indicated by the shaded purple area, used to compare $\eta$ and $\theta$ peaks in terms of $r$. This shaded region corresponds to the $\eta$-peak-to-peak $r$-range in \Cref{fig:table_FMINIST_MNIST_NIST_2000}(c).

Finally, note in \Cref{fig:tau_FMINIST_MNIST_NIST_2000} the slow down of diffusion due to the increasing ``volume'' of the network discussed in \Cref{sec:phasediagram} and indicated by the $\tau^{(2)}_{peak}$ curve as a function of $r$. We remark that $\tau^{(2)}_{peak}$ is a function of solely the parameter $r$. The latter shapes solely the degree distribution, as the number of nodes and edges is fixed across the phase plane and fixed in the regime of sparsity, symbolically $\delta(\frac{N-1}{|E|} - 1)$.

\subsection{Efficient performance at the emergence of scale-invariance}

In this and the next section, we discuss the relation between the location of the maximum pattern-matching efficiency on the algorithm's phase plane (\Cref{fig:table_FMINIST_MNIST_NIST_2000}(e--h)) and the dataset's complexity (\Cref{fig:dataset_complexity}).

All optimal networks are found in the amorphous region of the phase plane where the dilation symmetry is observed in the spectral density and reflected in the specific heat plateau. 
To discuss this observation we refer to the following argument~\cite{Cramer2020,Hidalgo2014,Goudarzi_2012}: a system needs tuning, or self-tuning, at the edge of a phase transition when the task demands a high level of complexity. Specifically, this was investigated in the context of second-order phase transitions~\cite{Cramer2020,Hidalgo2014}.
But what is the advantage of being close to a second-order phase transition? One possible advantage comes from the power-law behavior of the density function of some quantities, e.g. density of cluster sizes in percolation, or density of avalanche sizes and duration in systems with temporal dynamics. The power law provides the system with a large range of availability from the density of the quantity of interest, and the system may exploit such availability as a resource in the task, whether it is adaptation to a complex environment or other goal-oriented ones.  

In our case, the amorphous region of the phase plane, where the network traits display self-similar structures in a limited $\tau$-resolution range, exhibits the power law of the spectral density, governing the network frequencies.
Thus, better performances in terms of efficiency, particularly pattern-matching efficiency, can be explained in terms of network frequencies, which provide effective resources for the real-world disorder structuring task.

\subsection{Spectral dimension estimates the dataset complexity}
\label{sec:spectral_dim_dataset_complexity}

We now examine the location of the optimal $\theta$ efficiency in the amorphous region of the phase plane for each dataset. Thus, for each dataset, we inspect the traits of the highly efficient structures. 
We resort to the spectral dimension~\cite{Alexander1982,hattori1987gaussian,Burioni1996}, $d_s$, which generalizes the concept of Euclidean dimension to disordered structures. It takes non-integer values between 1 and 3 and the same value of the Euclidean dimension in the case of d-dimensional lattices.
The spectral dimension gauges the effective large-scale connectedness of disordered structures, with high values of $d_s$ indicating high topological connectedness, and is related to the asymptotic behavior of diffusion~\cite{hattori1987gaussian}, but also to electrical transport and thermal stability of folded proteins~\cite{Burioni2004}

Depending on where the maximal $\theta$-efficiency is observed in the amorphous region of the phase plane in \Cref{fig:table_FMINIST_MNIST_NIST_2000}, we see networks changing the value of the specific heat plateau, reflecting the spectral dimension, $d_s=2C_0$~\cite{Villegas_2022}, as we discussed in \Cref{sec:methods}. We also observe that the specific heat plateau is an approximately decreasing function of $r$ in the scale-invariant region of the phase plane in \Cref{fig:phase_diagram}(a). See also Sec. S4 in Supplemental Material~\cite{SM}.

For complex datasets, such as the CIFAR10 and FashionMNIST, the optimal $\theta$s are closer to the transition border delimiting the star graph from the amorphous phase. In contrast, for the simpler NIST and MNIST datasets, the optimal values are farther from their respective transition points. 
This observation also holds for the thermodynamic efficiency, $\eta$, noted the correspondence of the peaks of $\eta(\tau=10)$ and $\theta$ for these datasets in terms of $r$. 
A special case is again the MNIST, with its two-step transition in the amorphous region, where local maximum $\eta$ are at the first transition, $r\simeq0.35$, and the global maxima of $\theta$ and $\eta$ are at $r\simeq0.6$.

\begin{figure}[hbtp!]
\begin{center}
\includegraphics[width=1\columnwidth]{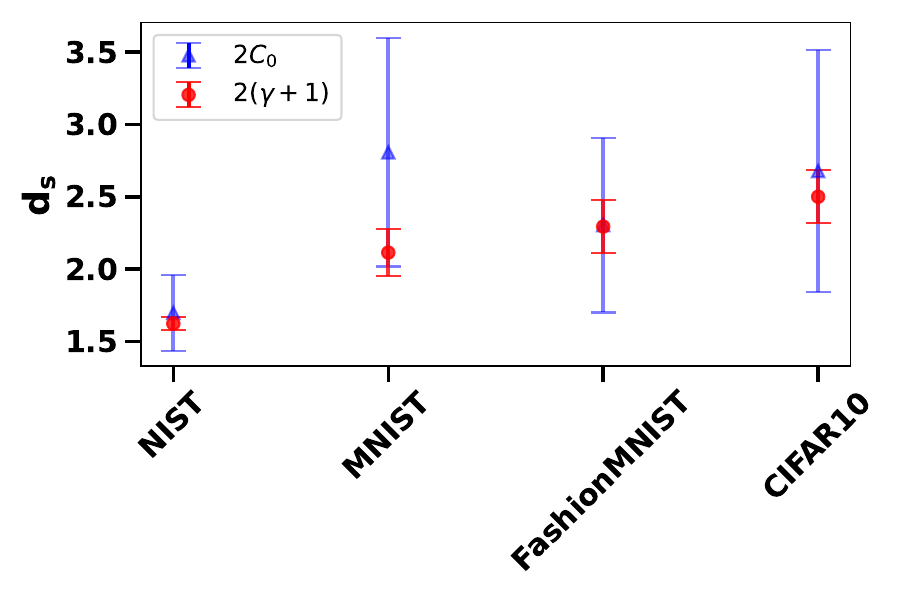}
\end{center}
\caption{\textbf{Datasets' spectral dimension}. The amorphous networks' spectral dimension is a good estimator of the dataset
complexity (\Cref{fig:dataset_complexity}). The red points indicate the spectral dimension~\cite{Burioni1996}, $d_s$, measured from the power-law exponent of the spectral density. The blue dots are obtained by monitoring the specific heat plateau~\cite{Villegas2023,poggialini2024networks}, $d_s=2C_0$, averaged over the resolution $\tau\in[1, 2]\times10^1$. All points are averaged over the $r$-range of the phase plane of maximal $\theta$-efficiency, see the shaded purple areas in \Cref{fig:table_FMINIST_MNIST_NIST_2000}(e--h), and over 50 independent simulations.}
\label{fig:dataset_spectral_dimension}
\end{figure}

We measure the spectral dimension for each dataset (N=2000) by averaging over the exponents of the spectral density $\mathcal{P}(\lambda)\sim\lambda^\gamma$. The exponents are obtained fitting the power law in the range $\lambda\in[1\times10^{-3}, 2\times10^{-1}]$ for all those networks which better adapted to the dataset, i.e. those included in the purple $r$-range of maximal $\theta$-efficiency in \Cref{fig:table_FMINIST_MNIST_NIST_2000}(e-h). The spectral dimension, $d_s=2(\gamma + 1)$~\cite{Burioni1996}, is plotted with red dots for each dataset in \Cref{fig:dataset_spectral_dimension}. 
We removed the value $\gamma(r=0.35)$ when measuring the spectral dimension of the CIFAR10 because the spectral distribution was too close to the transition to obtain a good fit of the power law.

We also estimated the spectral dimension by measuring the specific heat plateau~\cite{Villegas2023,poggialini2024networks}, $C_0$, over the resolution range $\tau\in[1, 2]\times10^1$. The measurements, $d_s=2C_0$, are shown with blue triangles in \Cref{fig:dataset_spectral_dimension}.  

\Cref{fig:dataset_spectral_dimension} shows that the amorphous networks' spectral dimension is a good estimator of the dataset complexity in \Cref{fig:dataset_complexity}. For higher-complexity datasets (ranked from left to right, from the low complexity NIST to the high complexity CIFAR10) we have increasing spectral dimension, $d_s$. 
In other words, the spectral dimension provides a way to gauge the complexity of the dataset after applying the tree generation protocol. Moreover, we can assess that high-complexity datasets are arranged in topological structures that are more compact when compared to those of lower complexity.

\section{Discussion}
\label{sec:discussion}
The ability of a system to survive is the capability of concentrating a flow of order from the environment - \textquote{drinking order from the environment}~\citep{Schrodinger2012}. 
In other words, systems, whether living systems or machine(-learning) systems, need to learn an internal representation of the complex world - or the dataset - they are posed in contact with. Ideally, to be competitive, such internal representation should be efficient in terms of desired function and available resources. It can be encoded in terms of abstract probability distributions~\cite{Mastromatteo2011,Hidalgo2014} or into the very network features of the system, as it happens in genetic regulatory networks playing a central role in morphogenesis~\cite{Krotov2014}. Or in a mixed way, as in artificial neural networks where the engineer designs a complicated distribution through the definition of the architecture and the training phase fills in the parameters to obtain the desired function-approximation machine able to achieve statistical generalization~\cite{Bengio2016}.

When confronted with the real-world environment, to survive, living systems not only need the capability of interpreting the environment, but they also need the capability to quickly react with a coherent response to stimuli, possibly drawn from the same distribution they adapted to~\cite{pryluk2019tradeoff}.
Moreover, the same system needs to cope with the cost of sending information across its structure, which can lead to an enhanced exploitation of resources.
In machine learning, the cost can be quantified by the number of computations required before the satisfying answer to a task, such as class matching, can be reported.

Here, we used an algorithm~\cite{Portegys} that creates and navigates trees of real-world patterns to investigate the intertwined role of structure and function in hierarchical self-organized tree systems in a simple setting.
The algorithm, akin to living systems, adapts to the diversity of the environment it is in contact with, represented here by the real-world dataset. It self-organizes structuring the disordered collection of high dimensional vectors into its network traits. The tree formation process undergoes two phase transitions at two different scales: the former is a mesoscale transition with the emergence of an approximately scale-invariant structural behavior progressively involving larger scales; the latter is a large-scale transition from the irregular scale-invariant structure to the regular-scale invariant 1d lattice.

Given that the original purpose of the algorithm was an efficient pattern-searching method, we have investigated the pattern-matching efficiency and its positioning on the phase-plane for four different datasets exhibiting distinct complexity levels. 

The recently developed theoretical tools of statistical physics for networks~\cite{De_Domenico_2016,Ghavasieh2020,Ghavasieh2023_generalized,Villegas_2022,Villegas2023,Ghavasieh2024} permit to quantify the diversity expressed in the traits of the network, with the von Neumann entropy, and the transmission flow of information, with the Helmholtz free energy. 
As the speed of information flow and sufficiently rich representation of the external world are two competing capabilities, the formation process of a complex structure can be interpreted as the macroscopic thermodynamic outcome of the delicate balance between the two. Thus, a variational principle gauged by the thermodynamic efficiency can be devised to explain why a certain structure emerges in place of another~\cite{Ghavasieh2024}.
In our work, this variational principle and the other tools provided by the theory are computationally employed to monitor the structural properties of sparse heterogeneous hierarchical trees produced by the Portegys protocol.

Empirical evidence suggests that the thermodynamic and the pattern-matching efficiencies show some consistency, namely when we look at the global maxima of the two in the $r$-range.
These two quantities involve two distinct processes that possess significant differences. On the one hand, the Portegys search phase is a directed process starting from the root node that navigates the structure under a homophily principle. On the other hand, the diffusion propagator describes an undirected ergodic process propagating from any possible node. 
Moreover, while the thermodynamic efficiency is defined based on general statistical physics arguments, and thus it is task-agnostic; the pattern-matching efficiency is defined based on the specific task under consideration. 

Still, we think it is reasonable to expect a similar behavior of the two efficiencies. Both quantities estimate a trade-off between the flow of a process and the encoding capabilities of the system's topological traits.
In both cases, the ``flow'' is quantified by a monotonic decreasing quantity as a function of $r$. For the Portegys we have the logarithmic number of computed distances, while for the Laplacian process, we have $\delta F = F + \frac{\log N}{\tau}$, where $F=-\frac{\log Z}{\tau}$. It is easy to see that these two quantities describe the flow in the network. The former counts the number of visited nodes across the traversing of the network by the Portegys search process. The latter is minus the logarithm of $Z$, with $Z\sim\mathcal{R}(\tau)$ where $\mathcal{R}(\tau)$ is the average return probability after time $\tau$ of a continuous time edge-centric random walker~\cite{Ghavasieh2020,ghavasieh2020enhancing}. Thus even though they are not alike, they can still be used as proxies of the ``flow''.

We have discussed the denominator of the two efficiencies: $\eta$ and $\theta$. We now discuss the numerator.
For what concerns the Portegys algorithm, we have observed that there are more descriptive network structures \textit{per se}. The accuracy is not necessarily a monotonic decreasing function of the control parameter, $r$. It can exhibit higher values at specific $r$-values as illustrated by the cases of the CIFAR10 and the FashionMNIST.

The matter is what makes a certain structure better suited in terms of an objective, such as accuracy or efficiency, to a given task. Equivalently and more generally said, in terms of a complex input-output mapping. When dealing with artificial neural networks, we can either enlarge the number of parameters or adjust their values to achieve more complicated functions, even though neither implies a one-to-one relationship with better generalization performance. In the case of our network structures, the parameters are the network traits themselves, which in the Fourier space become the ``network frequencies'', $\lambda_i$, i.e. the Laplacian eigenvalues. We can measure a weighted average of the parameters by computing the thermal average of the frequencies contained in the network, precisely the numerator in \Cref{eq:efficiency}. The internal energy, $U$, is the Gibbs average of the Laplacian eigenvalues of the network at the operational resolution scale, $\tau$, tantamount to the inverse temperature, $\beta$, in statistical physics. Thus, the Gibbs average frequency content is quantified by the internal energy, $U_\tau=\Tr[\hat{\rho}\hat{L}] = <\lambda>_\tau$, where for scale-invariant networks $<\lambda>_\tau = C_0 / \tau=\frac{1}{2}\frac{d_s}{\tau}$~\cite{Villegas2023} in a limited $\tau$ range due to the finite size of network and the frequency cut-off that we discuss in the following. Thus, $U$ is related to the specific heat plateau, $C_0$, and, importantly, to the spectral dimension~\cite{Burioni1996}, $d_s$, which governs the spectral density. The difference to the case of the artificial neural network is that we can’t directly increase the number of parameters, given a fixed number of samples, or directly adjust the parameter values. Still, we can govern the distribution of frequencies with the control parameter, $r$, and let the structure self-organize. 

We have computed the thermodynamic efficiency at a specific resolution scale, $\tau=10$, identified by inspecting the specific heat, $C=-dS/ d \log \tau$. At this $\tau$-scale, we could monitor the transition from the trivial structure, a single specific heat peak, to the emergence of complex amorphous structures, where the specific heat shows a plateau. The plateau reflects the growth of an approximately scale-invariant regime where information is integrated at a constant rate. Such $\tau$-resolution scale is akin to an ``ultra-violet'' cut-off, delimiting from above the relevant frequencies of the networks across the transition to scale-invariance. At the same time, it scans the scale where modules emerge in the (loop-less) network. For this reason, we could gauge the phase transition to the amorphous regime with the network modularity, $Q$, as an order parameter. Finally, this resolution $\tau$-scale is independent of the control parameter, the dataset, and the doubling of the number of nodes (see Supplemental Material~\cite{SM}).

Highly efficient networks were always found in the amorphous regime. From a functional point of view, we mention that several works in the literature~\cite{Cramer2020,Hidalgo2014,Goudarzi_2012} suggest that a system benefits from the vicinity to a phase transition critical point when it confronts environments of high complexity, while it does not benefit from it when the environmental complexity is lower. This partly aligns with our findings on the four datasets of increasing complexity. One key aspect of being close to a critical point is to take advantage of the functional benefits that a power law density related to some quantity of interest has to offer. In our case, for all datasets, we saw higher performance in terms of pattern-matching efficiency in the amorphous region of the phase plane, characterized by a power law spectral density.

Nevertheless, when the dataset was more difficult, an optimal efficiency to the pattern-searching task was found close to the transition border separating the star graph and the scale-invariant regime. On the contrary, lower complexity datasets find their highly efficient structures on the opposite side of the amorphous phase plane region. 

Then what is the reason for a network embedding high dimensional patterns to be in one or the other location of the phase plane? 
By measuring the optimally efficient networks' spectral dimension~\cite{Alexander1982,Burioni1996}, $d_s$, almost continuously changing in the amorphous regime, we could relate the dimensionality of the network to the complexity of the dataset it structured. 
We observed higher spectral dimension, reflecting large-scale connectedness, for networks mapping high-complexity datasets. On the contrary, easier datasets were mapped into more complex branching structures with a lower spectral dimension. Thus, importantly, we provide an alternative way to measure the complexity of a dataset: by mapping it into a tree network and then measuring its spectral dimension.

Returning to the starting question of this manuscript—why a certain structure emerges in place of another—the answer suggested by our findings is: to manage the complexity of the environment it is adapting to. 

The generalization of this possible answer to systems other than the Portegys algorithm remains purely speculative and beyond the scope of this work. Still, the methods presented can in principle provide diagnostics to other tree-based approaches to classification such as hierarchical clustering. 

Suppose we are given a pre-trained tree network mapping an unknown dataset, we can estimate the complexity of the dataset that was used for training, along with the ability of the tree structure to cope with that complexity without performing any test phase.
Suppose the trained network ``struggles'' in the representation of the dataset. In such a case, the spectral dimension is expected to be high, and the network is expected to exhibit higher connectedness. On the contrary, a network mapping a low-complexity dataset should exhibit branched structures and low spectral dimension.

In the reverse direction, knowing the complexity of a dataset, we could initialize the search for the optimal control parameter, $r$, closer to values typically presenting one of the two transitions exhibited by the algorithm, reducing the number of simulations required to find the most efficient network structure.

Interestingly, developments of our analysis and methods on hierarchical tree networks could be applied to other topics in artificial intelligence, neuroscience, and neuromorphic computing. 

We find it particularly suited to the context of biologically inspired~\cite{Poirazi2020,Hodassman2022,Shepherd1987} dendritic learning~\cite{Chavlis2025,Guerguiev2017,Payeur2021,Sacramento2018}, where input units are connected to output units via loop-less paths, to guide the design of the tree-like dendritic structures for parameter-efficient learning in artificial neural network~\cite{Chavlis2025,Guerguiev2017,Payeur2021,Sacramento2018}.

Neurons can be modeled as the amorphous tree structures discussed in this work. 
It would be interesting to experimentally relate the dendritic structure, specifically its spectral dimension, to the capability of the neuron to deal with environmental stimuli and complicated input-output mapping of signals~\cite{Shepherd1987,Poirazi2020,Hodassman2022,Kastellakis2016,Makarov2023}, or to produce signals with richer diversity~\cite{Mainen1996,pryluk2019tradeoff,vanElburg2010}.
For instance, a neuron capable of dealing with the complexity of the stimulus it is subject to would be expected to exhibit traits with low spectral dimensionality and complex branching.

Moreover, potential applications include the topic of neuromorphic computing~\citep{Mead2020,Indiveri_2013,Christensen_2022} currently looking for a possible solution to the ever-increasing computational power demand of recent years. Neuromorphic computing processors are designed to co-localize computation and memory on the physical substrate and emulate the brain learning capabilities \textit{in materia}. Complete neural networks with dendritic branching have been fabricated with memristive devices and have shown significant power efficiency~\cite{Li2020}.
The full potential of biological neural systems is likely achieved because of the interplay between their complex topological structures and the functional dynamics of a very large number of elements evolving on it~\citep{Suarez_2021}. 
Analogously to biological neural systems, structurally efficient routing may play a key role for \textit{in-materia} computing processors fabricated with self-assembled materials~\cite{Vahl2024,FusedMembrain}. Emergent tree-like structures are currently physically fabricated in emerging materials representing potential candidates for neuromorphic computing. We mention tree structures produced at the critical point of a percolation transition for randomly dispersed nanoparticles~\cite{Lee2015,Mallinson2019}, or the electrochemically formed dendrite morphology at the origin of the resistive-switching of some memristive devices~\cite{Lee2015,Guo2007}, and fractal trees produced in the dielectric breakdown~\cite{Niemeyer1984}. Thus, their information processing capabilities could be investigated and enhanced based on this work and future developments.

\section{Conclusions}
 
We summarize the contributions included in this work.

(i) We proposed the Portegys algorithm~\cite{Portegys,Mangalagiu1999} as an alternative process to produce hierarchies for statistical physics analysis, integrating the functional aspect of a machine-learning algorithm.

(ii) Multiple phase transitions are identified while the algorithm searches for an efficient internal representation of the dataset.

(iii) Scale-invariance, i.e. power-law Laplacian spectral density~\cite{poggialini2024networks}, is a key feature to constructing efficient tree-networks, capable of combining fast information flow and sufficiently rich internal representation of information. Our analysis involved task-agnostic thermodynamic-like efficiency~\cite{Ghavasieh2024}, based on general statistical physics arguments, and task-related pattern-matching efficiency, quantifying the trade-off between accuracy and resources used to achieve the task.

(iv) Studying efficient tree structures mapping the structured disorder of a dataset, we showed that the emergence of complex trees is directly influenced by the complexity of the environmental conditions. Namely, the spectral dimension~\cite{Alexander1982,hattori1987gaussian,Burioni1996}, proportional to the exponent of the power-law spectral density, directly encodes the complexity of the disordered environment.  

(v) Therefore, we proposed two novel operational measurements to estimate the dataset complexity: by measuring the spectral dimension of the highly efficient network mapping the dataset; or, equivalently, by measuring the specific heat plateau~\cite{Villegas_2022,Villegas2023}, which reflects the spectral dilation symmetry and the network resolution where information is integrated at a constant rate.

\section*{Conflict of Interest Statement}
The authors declare that the research was conducted in the absence of any commercial or financial relationships that could be construed as a potential conflict of interest.

\section*{Author Contributions}
DC: Theoretical framework, Conceptualization, Methodology, Simulations, Data curation, Software, Writing – original draft, Writing – review \& editing. LS: Writing – review \& editing, Conceptualization, Supervision, Funding acquisition.

\section*{Funding}
This work was funded by EUs Horizon 2020, from the MSCA-ITN-2019 Innovative Training Networks program ``Materials for Neuromorphic Circuits" (MANIC) under the grant agreement No. 861153, as well as by the financial support of the CogniGron research center and the Ubbo Emmius Funds (Univ. of Groningen).\newline

\section*{Acknowledgments}
We are thankful to Jos van Goor for providing some of the scripts that enabled efficient grid search across parameters and to Federico Balducci for the useful discussions.


\section*{Data Availability Statement}
The original contributions presented in the study are included in the article/supplemental material, further inquiries can be directed to the corresponding author/s. Datasets and scripts can be found at \url{https://github.com/CipolliniDavide/Laplacian_Trees}.


\bibliography{references.bib}



\end{document}